\documentclass{article}
\usepackage[left=2cm,right=2cm]{geometry}
\usepackage{graphicx}
\usepackage{amsmath}
\usepackage{amssymb}
\usepackage{float}
\usepackage{graphicx}
\usepackage{subcaption}

\def\tstau{{\tilde \tau}'}
\def\stau{{\tilde \tau}}
\def\msusy{m_{\mathrm{SUSY}}}
\begin{document}
	\title{Twin stau as a self-interacting dark matter
	}
	\author{
		Micha\l { } \L ukawski        \\
		{\em Institute of Theoretical Physics, Faculty of Physics, University of Warsaw,} \\ {ul. Pasteura 5, PL-02-093 Warsaw, Poland } \\
		{\em mlukawski@fuw.edu.pl}
	}
	\maketitle
	\baselineskip=10pt
	\begin{abstract}
		Supersymmetric Twin Higgs models allow for reducing the fine-tuning with respect to Minimal Supersymmetric Standard Model by protecting the mass of the Higgs boson by additional, accidental global symmetry. This class of models introduce numerous new states, some of which might be candidates for dark matter. Since those reside in twin sector, they are not charged under Standard Model gauge group. We proposed twin stau as a candidate for dark matter. Even though twin stau is charged under twin electromagnetism, since supersymmetric partners obtain large masses from supersymmetry breaking they can easily escape bounds for self-interacting dark matter. The mass of twin stau which reproduces correct relic abundance is usually between 300 and 500 GeV. This scenario can be probed by future direct detection experiments such as Lux-Zepelin.
	\end{abstract}
	\baselineskip=14pt

	\section{Introduction}
	The mass of the Higgs particle is not protected from the large corrections from higher energy scales such as Planck mass. One of the most successful mechanisms protecting the mass of the Higgs is supersymmetry, which allows for cancellation of the these corrections above the scale of supersymmetry breaking, $\msusy$, at which new particles enter \cite{Kaul:1981wp}. Naturalness requires that $\msusy$ is not much above the EW scale, usually at below 1 TeV. Hence, one of the main features of supersymmetric models are relatively light coloured particles which should be produced at LHC in hadron collisions. Since no such signal has been found \cite{ATLAS:2020dsf}, the minimal implementations of supersymmetry such as Minimal Supersymmetric Standard Model (MSSM) requires large fine-tuning of the parameters to reproduce the correct electroweak scale. \\
	One of the mechanisms which can relax the fine-tuning of supersymmetric models is Twin Higgs (TH), first introduced in non-supersymmetric context \cite{Chacko:2005pe}. TH mechanism extends the particle content with partners of all MSSM states and imposes $Z_2$ symmetry between sectors, which we refer to as visible (even though supersymmetric particles have not been observed yet) and twin. The Higgs particle in this class of models is then the pseudo-Nambu-Goldstone boson (pNGB) of an accidental, global symmetry of the potential. Its mass is generated by explicit breaking of the global symmetry. Since the mass of the Higgs is proportional to small breaking of the symmetry, it is protected from large quantum corrections. \\
	Twin Higgs models predict existence of numerous new states, some of which could be dark matter (DM) candidates, including dark matter in
	twin supersymmetric sector\cite{Badziak:2019zys}. However, in cases of charged DM, one has to either break or eliminate twin electromagnetism gauge symmetry to escape bounds on self-interacting dark matter. In the following, I will show that in Supersymmetric Twin Higgs (SUSY TH) models twin stau is a viable DM candidate with interesting properties such as long range self-interactions mediated by the twin photon. \\
	More detailed discussion of the results presented here can be found
	in \cite{Badziak:2022eag}.

	\section{Supersymmetric Twin Higgs}
	First, let's take a look at the scalar potential of a TH model without referring to supersymmetry for simplicity. As mentioned before,  the particle content of the Standard Model is doubled by adding a second, twin sector. In particular, the scalar sector is extended by twin Higgs doublet $H'$. Additionally, $Z_2$ symmetry interchanging particles between sectors is imposed. The potential is given by \cite{Craig:2015pha}
	\begin{equation}
		V(H,H')= \lambda \big( H^2+H'^2\big)^2 - m_\mathcal{H}^2\big( H^2+H'^2)+ \Delta \lambda \big( H^4+H'^4 \big) +\Delta m^2 H^2
		\label{SU4potential}
	\end{equation}
	Note that these doublets form $SU(4)$ fundamental representation $\mathcal{H}_{\mathrm{SU(4)}}=(H,H')$. Having that in mind, we clearly see that first and second terms are $SU(4)$ and $Z_2$ invariant. These terms form Mexican hat potential, which leads to the spontaneous symmetry breaking. The third term breaks $SU(4)$ while preserving $Z_2$ and is responsible for generation of the mass of  the Higgs particle. The last term breaks $Z_2$, generating misalignment of the vacuum expectation values (vevs) between sectors parameterized by ratio $v'/v$. It is necessary since otherwise exact $Z_2$ symmetry would imply equal decay branching ratio of the Higgs into visible and twin sectors, which is in contradiction with LHC data. Current constraint on invisible Higgs decays imply that $v'/v \gtrsim
	3$, \cite{Craig:2015pha}. Misalignment of vevs requires fine-tuning of parameters needed to reproduce the mass of the visible Higgs, which can be parameterized by $\Delta_{v'/v}=(v'^2/v^2-2)/2$ which leads to $\mathcal{O}(30)$\% fine-tuning for $v'/v=3$ which will be our benchmark for discussion. Fine-tuning at that level means that model is essentially fully natural. It should be mentioned that ratios $v'/v\geq 7$ are disfavored since they require fine-tuning worse than 5\%. \\
	As mentioned before, the minimum of the potential breaks symmetry $SU(4)$ to $SU(3)$ generating 7 pNGBs. Six of them give masses to $SU_L(2)$ and $SU_L'(2)$ gauge bosons, while the remaining one is identified with the SM Higgs. \\
	In supersymmetric models, the scalar potential is fully fixed by the gauge interactions (D-terms) and the particle content (F-terms). Hence, the scalar potential of the form (\ref{SU4potential}) cannot be simply added to the lagrangian and should come from either D-term \cite{Badziak:2017kjk,Badziak:2017syq,Badziak:2017wxn} or F-term \cite{Falkowski2006,Chang:2006ra}. In both cases, the tree-level mass of the Higgs is proportional to $\cos^2(2\beta)$, which approaches 1 in the limit of large $\tan\beta= v_u/v_d$. The key difference between those cases is captured by $\lambda$ dependence on $\tan\beta$. In general, the fine-tuning in SUSY TH is inversely proportional to the quartic coupling $\lambda$ of the $SU(4)$ invariant term of the potential.  \\
	In F-term SUSY TH, quartic term is proportional to $\sin^2(2\beta)$, which is maximized for $\tan\beta=1$ for which tree-level mass of the Higgs vanishes. It is then necessary to use moderate values of $\tan\beta$, which inevitably lead to fine-tuning which requires further model building to alleviate. \\
	In contrast, in D-term SUSY TH, the quartic term is proportional to $\cos^2(2\beta)$ and is maximized for large values of $\tan\beta$. As a result, fine-tuning in these models can be as low as 20\% with heavy stops masses of 2 TeV. We will focus on this case since it seems far more preferable, however most of the analysis presented here is not UV dependent.
	
	\section{Twin stau}
	Twin stau is $Z_2$ partner of the supersymmetric scalar partner of tau lepton. Hence, it is charged under twin electromagnetism and twin weak interactions $U_Y'(1)\times SU_L'(2)$ and its mass gets large contributions from the supersymmetry breaking. We will consider $Z_2$ symmetric SUSY breaking, thus soft masses twin stau obtains are equal in visible and twin sectors. To be more specific, the mass matrix of twin stau is given by \cite{Martin:1997ns} \begin{equation}
		m_{\Tilde{\tau'}}^2 = \left( \begin{matrix} m_{L_3}^2 + \Delta_{\Tilde{\tau}_L} +m_{\tau'}^2 & -\mu v' y_\tau \sin (\beta) \\-\mu v' y_\tau \sin (\beta))  &  m_{\Bar{{e}}_3}^2 + \Delta_{\Tilde{\tau}_R} +m_{\tau'}^2 \end{matrix} \right)
	\end{equation}
	where $m_{L_3}$ and $m_{R_3}$ are soft SUSY breaking masses, $\mu$ is Higgs mass term, $v'$ is vacuum expectation value of twin Higgs and $m_{\tau'}$ is mass of the twin tau. Note, that it is assumed that there is no $Z_2$ breaking in Yukawa sector $y_{\tau'}=y_{\tau}$ and $\tan\beta'=\tan\beta$. The D-term contributions to the mass of the twin stau are given by $\Delta_{\Tilde{\tau}_L'}=(-1/2+\sin^2\theta_W)\cos2\beta m_Z'^2$ and $\Delta_{\Tilde{\tau}_R'}=-\sin^2\theta_W\cos2\beta m_Z'^2$, where $\theta_W$ is Weinberg angle. Mass matrix of stau is obtained by removing all primes. \\
	Note that for $v'/v>1$, the off-diagonal terms are larger in twin sector than in visible one. It leads to twin stau lighter than stau in the parameter space with large mixing. By making a common assumption that lightest supersymmetric particle is stable, the twin stau can be candidate for a dark matter. \\
	Note that while twin stau DM prefers large off-diagonal mass matrix, $\mu$ cannot be arbitrarily large if the model is to remain natural. Assumption of equal Yukawa couplings in both sectors is not necessary and might even be preferable (see \cite{Chacko:2016hvu} for $\Delta N_{\mathrm{eff}}$ problem in TH models and how it could be solved by breaking $Z_2$ in Yukawas\cite{Chacko:2016hvu,Barbieri:2017opf}), however I will not cover this case. \\
	As mentioned in the introduction, twin stau is charged under twin electromagnetism, which implies that there are long-range self-interactions mediated by massless twin photons. As discussed in \cite{Agrawal:2016quu} the bound on mass of self-interacting DM have been overestimated and currently the strongest constraint comes from measurements of non-zero ellipticity of gravitational potential of NGC720. For equal couplings of electromagnetism and twin electromagnetism, the lower bound on mass of twin stau is approximately 210 GeV. That is the reason why breaking of twin electromagnetism is necessary in case of charged, non-supersymmetric candidates for dark matter, such as twin taus or mesons.
	
	\subsection{Direct detection}
\begin{figure}[htb]
		\begin{center}
			{\includegraphics[scale=0.3]{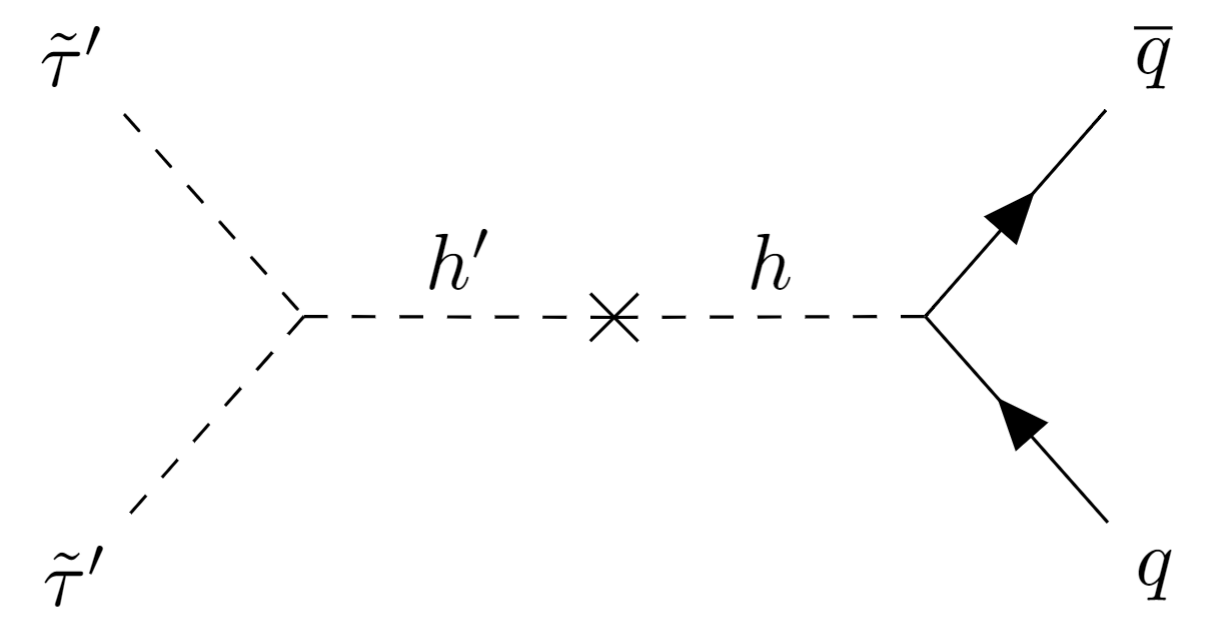}}\hspace{0.5cm}
			\caption{Diagram of twin stau interactions with visible sector quarks. The interaction is mediated by the Higgs portal}
			\label{DD}
		\end{center}
	\end{figure}

	Since twin stau belongs to the twin sector, its only interactions with the visible sector must be mediated by the Higgs portal, Fig. \ref{DD}. Thus, the interactions of twin stau with atomic nuclei are suppressed by mixing between Higgs and twin Higgs, which is roughly given by $v/v'$. In particular, the effective interaction between the visible SM Higgs and twin stau is given by
	\begin{equation}
		\lambda_{h\tilde{\tau}'\tilde{\tau}'}= \frac{g}{m_{W'}}\left[ \left(\frac{1}{2} c_{\theta_{\tilde{\tau}'}}^2-s_W^2 c_{2\theta_{\tilde{\tau}'}}\right) m_{Z'}^2 c_{2\beta} - m_{\tau'}^2 + \frac{m_{\tau'}}{2} \mu \tan\beta s_{2\theta_{\tilde{\tau}'}}  \right]   \frac{v}{v'}
	\end{equation}
	where $c_\alpha$ and $s_\alpha$ are shortcuts for $\cos\alpha$ and $\sin\alpha$, respectively, and $m_{W'}$ is twin W boson mass. Note that this coupling is maximized for large mixing angle $\theta_{\tilde{\tau}'}$, so we expect DD bounds to be strongest in that region. \\
	In our results, we will assume that in the region with $\Omega h^2>0.12$, there exists a mechanism which dilutes DM to observed relic abundance. For region with $\Omega h^2<0.12$, we rescale the DM cross section with factor $\Omega h^2/0.12$. It is justified by the fact that it is possible that twin stau constitutes a fraction of DM, but overproduced DM is not physical. A more detailed discussion of the issue can be found in \cite{Badziak:2022eag}.
	
	\section{Results}
	For the calculation of the relic density, we have modified \texttt{Micromegas} \cite{Belanger:2001fz,Belanger:2004yn,Belanger:2006is}. \texttt{Micromegas} takes into account twin stau coannihilations within a twin sector, but since in a large portion of the parameter space stau is almost degenerate with twin stau it is necessary to adjust the relic abundance. Upon justified assumption that the annihilation cross section for stau and twin stau is the same, the effective cross section for twin stau is given by \begin{equation}
		\sigma_{\mathrm{eff}} =\sigma \frac{1+(1+\Delta)^3 e^{-2 x_f \Delta}}{[1+(1+\Delta)^{3/2} e^{-x_f \Delta}]^2},
	\end{equation}
	where $x_f=m_{\tstau}/T_f\sim25$, with $T_f$ the freeze-out temperature. The relic density can then be approximated using $\Omega_{\mathrm{coann}}h^2=\Omega_0 h^2 \sigma/\sigma_{\mathrm{eff}}$.

	\begin{figure}[h]
		%\centering
		\begin{subfigure}{.5\textwidth}
			%\centering
			\includegraphics[scale=0.9]{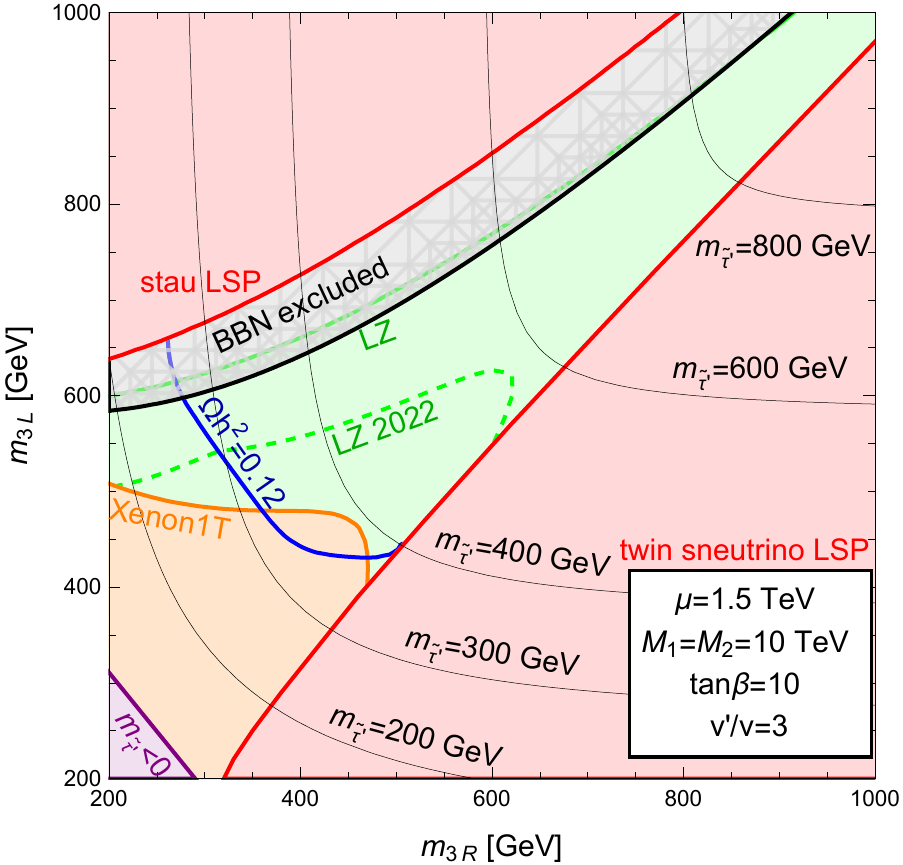}
			\label{fig:sub1}
		\end{subfigure}%
		\begin{subfigure}{.5\textwidth}
		%	\centering
			\includegraphics[scale=0.9]{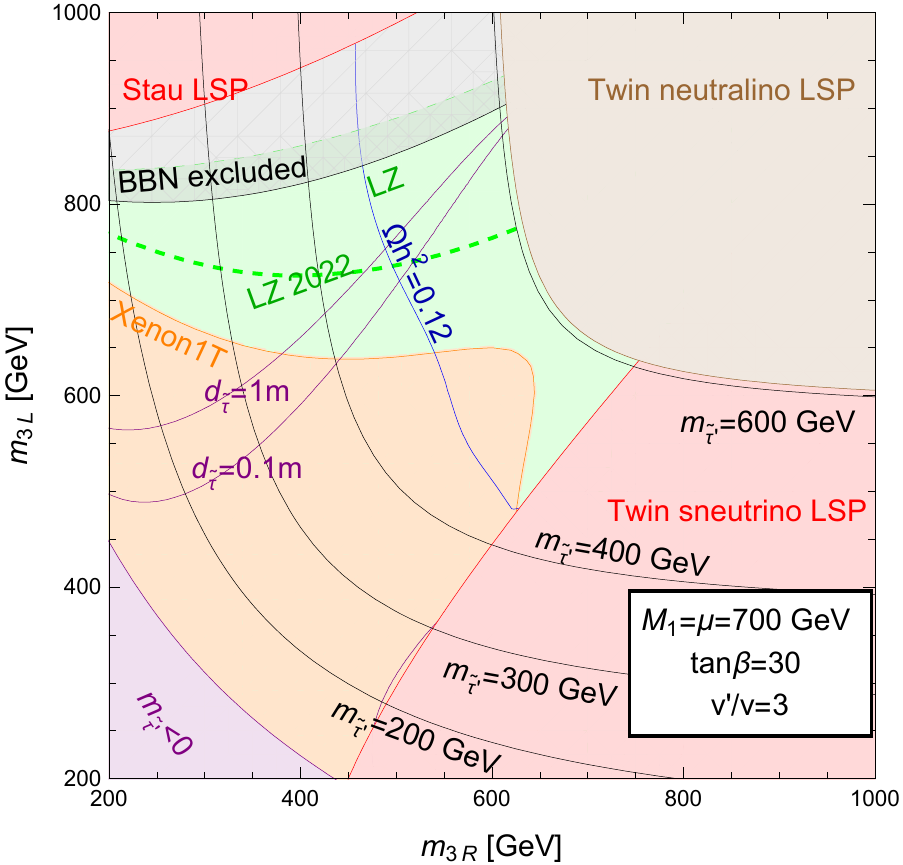}
			\label{fig:sub2}
		\end{subfigure}
		\caption{\it Contour of relic abundance $\Omega h^2=0.12$ (blue line) in plane of soft twin stau masses $m_{R3}$ and $m_{L3}$. In purple region twin stau is tachyonic, in red twin stau is not LSP. Mass contours of twin stau are black.  Direct detection bounds from Xenon1T and LZ are coloured orange and green, respectively. Dashed green contour corresponds to new, first results from LZ \cite{LZ:2022ufs}}
		\label{mRmL}
	\end{figure}
	
	First, we will consider the decoupling case where all SUSY breaking masses except the stau are set to 10 TeV, left panel of Fig. \ref{mRmL}. We consider $\mu=1.5$ TeV which is unnatural but is a good starting point for the analysis. As mentioned before, large $\tan\beta$ is preferred both due to naturalness and large twin stau mixing necessary for twin stau LSP. Some of the parameter space is excluded due to mass spectrum. For mostly left-handed twin stau twin sneutrino is always LSP while for mostly right-handed $\tstau$ stau is the LSP and is excluded. For very small soft masses, twin stau is tachyonic due to off-diagonal mass terms proportional to the large $\mu$. In this case, the correct relic abundance is obtained for $m_{\tstau}$ between 260 and 400 GeV, depending on the mixing. Some of the parameter space is excluded due to the direct detection bounds coming from Xenon1T \cite{XENON:2018voc} and primary results from Lux-Zepelin (LZ) \cite{LZ:2022ufs}. Predicted sensitivity of LZ will allow for probing whole parameters space shown on this plot. The Big Bang Nucleosynthesis (BBN) bound comes from the fact that in this region of parameter space, the difference between masses of stau and twin stau becomes too small to allow $\stau\rightarrow\tstau^\dagger\tau\tau'$ decay. As a result, stau becomes long-lived, and its late decay would change the nuclei composition of the universe. However, in the decoupled case, the lifetime of stau is generically too long. Charged long-lived particles (with decay length above $\mathcal{O}(1)$m) could have been seen at LHC as charged disappearing tracks if their mass is lower than approximately 430 GeV \cite{ATLAS:2018lob}. Since stau decay is mediated by twin bino and higgsino big, $\mu$ and very large $M_1$ lead to large decay length across whole plot. \\

	A more realistic scenario with $M_1=\mu=700$ GeV is shown on right panel of Fig. \ref{mRmL}. A large portion of that parameter space has decay length of stau below 1 m. Note that in this plot only small part is not excluded by either new results from LZ nor by decay length of stau. However, the reason for that is that we keep the most natural, non-excluded value of vevs ratio $v'/v=3$. One can trade off naturalness for the opening up of the parameter space. I decided to keep the ratio $v'/v=3$ as originally presented at \texttt{7th Young Researchers' Workshop}, as well as in \cite{Badziak:2022eag}. However, one must keep in mind that new LZ results indicate that the scenario with minimal tuning is now strongly constrained. The final results from Lux-Zepelin will probe the whole parameters space.
	
	\section*{Acknowledgements}
	This work was partially supported by the
	National Science Centre, Poland, under research grant no.
	2020/38/E/ST2/00243.

\end{document}